\documentclass[letter]{aa}  

\usepackage{array}
\usepackage{times}
\usepackage{mathptmx}
\usepackage[T1]{fontenc}
\usepackage{ae,aecompl}

\usepackage{amsmath}    % Advanced maths commands
\usepackage{amssymb}
\usepackage{bm}
\usepackage{graphicx}
\usepackage{xcolor}
\usepackage{txfonts}
\usepackage{hyperref}
\begin{document}

   \title{First observations and magnitude measurement of Starlink's Darksat}
\titlerunning{Observations of Starlink's Darksat}

   \author{J. Tregloan-Reed\inst{1}
          \and             
          A. Otarola\inst{2}
          \and
          E. Ortiz\inst{3}
          \and
          V. Molina\inst{1}
          \and
          J. Anais\inst{1}
          \and
          R. Gonz\'{a}lez\inst{1}
          \and \\
          J. P. Colque\inst{1}
          \and
          E. Unda-Sanzana\inst{1}
          }
\authorrunning{Tregloan-Reed et al.
}

   \institute{Centro de Astronom\'{i}a (CITEVA), Universidad de Antofagasta,
              Avenida U. de Antofagasta 02800, Antofagasta, Chile\\
              \email{jeremy.tregloanreed@uantof.cl}
              \and
              TMT International Observatory, 100 West Walnut Street, Pasadena, CA 91124, USA.
              \and
              Departamento de F\'{i}sica, Universidad de Antofagasta, Avenida Angamos 601, Antofagasta, Chile
             }

   \date{Received March 13,2020; accepted April 15,2020}

% \abstract{}{}{}{}{} 
% 5 {} token are mandatory
 
  \abstract
  % context heading (optional)
  % {} leave it empty if necessary 
  {}
  % aims heading (mandatory)
   {We measured the Sloan g' magnitudes of the Starlink's STARLINK-1130 (Darksat) and 1113 LEO communication satellites to determine the effectiveness of the Darksat darkening treatment at 475.4\,nm. }
  % methods heading (mandatory)
   {Two observations of the Starlink's Darksat low Earth orbit (LEO) communication satellite were conducted on 2020/02/08 and 2020/03/06 using Sloan r' and g' filters, respectively. A second satellite, STARLINK-1113, was observed on 2020/03/06 using a Sloan g' filter. The initial observation on 2020/02/08 was a test observation conducted when Darksat was still in the process of manoeuvring to its nominal orbit and orientation. Based on the successful test observation, the first main observation took place on 2020/03/06, along with an observation of the second Starlink satellite. }
  % results heading (mandatory)
   {The calibration, image processing, and analysis of the Darksat Sloan g' image gives an estimated Sloan g' magnitude of $\mathbf{7.46\pm0.04}$ at a range of 976.50\,km. For STARLINK-1113, an estimated Sloan g' magnitude of $\mathbf{6.59\pm0.05}$ at a range of 941.62\,km was found. When scaled to a range of 550\,km and corrected for the solar and observer phase angles, a reduction by a factor of two is seen in the reflected solar flux between Darksat and STARLINK-1113. }
  % conclusions heading (optional), leave it empty if necessary 
   {The data and results presented in this work demonstrate that the special darkening coating used by Starlink for Darksat has darkened the Sloan g' magnitude by $0.77\pm0.05$\,mag when the range is equal to a nominal orbital height (550\,km). This result will serve members of the astronomical community who are actively modelling the satellite mega-constellations to ascertain their actual impact on both amateur and professional  astronomical observations. Both concurrent and subsequent observations are planned to cover the full optical and NIR spectrum using an ensemble of instruments, telescopes, and observatories. }


   \maketitle
%
%-------------------------------------------------------------------

\section{Introduction}\label{Sec:1}

In May 2019, Starlink, a subsidiary of SpaceX, launched their first batch of 60 low Earth orbit (LEO) communication satellites. Due to the extremely bright apparent magnitude of these satellites, caused by their very low orbits following launch and their clustering in `trains', this launch and subsequent launches have caused major concern among both the amateur and professional astronomical communities (see \href{https://www.iau.org/news/announcements/detail/ann19035/}{IAU press release, 2019/06/03} and \href{https://www.iau.org/news/pressreleases/detail/iau2001/}{IAU press release, 2020/02/12}).  \protect\citet{Hainaut2020} recently examined the impact on ESO telescopes in the optical and NIR and suggest that the greatest impact from the LEO communication satellites will be on ultra-wide imaging exposures from large telescopes (e.g. National Science Foundation’s Vera C. Rubin Observatory, formerly known as LSST). A second study, carried out by  \citet{McDowell2020}, concludes that LEO satellites will have a significant impact on twilight astronomy and observations that use wide fields of view with long exposures.

Starlink, via an application submitted by SpaceX, has received approval by the U.S. FCC (Federal Communications Commission) to have 12\,000 Starlink LEO communication satellites in orbit (\href{https://docs.fcc.gov/public/attachments/DA-19-342A1.pdf}{FCC Authorization Report} and \href{https://docs.fcc.gov/public/attachments/DOC-355102A1.pdf}{FCC statement}). SpaceX have since applied to the international radio-frequency regulator for a further 30\,000 Starlink LEO communication satellites to be placed in low orbit\footnote{\href{https://spacenews.com/spacex-submits-paperwork-for-30000-more-starlink-satellites}{Spacenews.com} accessed on 2020/03/01} (328\,km to 580\,km). The threat to ground-based optical and radio astronomical research from the Starlink mega-constellations is still being evaluated by the AAS and IAU. Starlink is working with with the astronomical community to reduce the brightness of the Starlink satellites\footnote{\href{https://www.theatlantic.com/science/archive/2020/02/spacex-starlink-astronomy/606169/}{The Atlantic} accessed on 2020/02/07}. 

Following these discussions, Starlink undertook an attempt to make the satellites dark enough so that they do not saturate the Charles Simonyi telescope's camera detectors at the Vera C. Rubin Observatory. At 02:19\,UTC on January 7 2020, Starlink launched its third batch of LEO communication satellites and one of them, STARLINK-1130 (nicknamed `Darksat') was given an experimental darkening treatment on one side to reduce its reflective brightness (SpaceX press kit, \href{https://www.spacex.com/sites/spacex/files/starlink_media_kit_jan2020.pdf}{January 2020}), although the exact details of this treatment have not been published.

In this letter, we present the first ground-based observations of the Darksat (international designation: 44932) LEO communication satellite (observed on 2020/02/08 and 2020/03/06)\footnote{The FITS files are available from the author by request.}, along with an additional non-darkened Starlink LEO communication satellite, STARLINK-1113 (international designation: 44926) (observed on 2020/03/06), as a comparison.

%--------------------------------------------------------------------
\section{Observations}\label{Sec:2}

Two observations of Darksat (STARLINK-1130) were conducted on 2020/02/08 and 2020/03/06, whereas STARLINK-1113 was observed on 2020/03/06, using the Chakana 0.6\,m telescope at Universidad de Antofagasta’s Ckoirama observatory in northern Chile (24.1\,S, 69.9\,W). The instrument  was a FLI ProLine 16801 camera, operated with a Sloan r'(2020/02/08) and g' (2020/03/06) filter. In this setup, the CCD covers a field of view of $32.4\times 32.4$\,arcmin with a pixel scale of 0.47 arcsec pixel$^{-1}$. 

Our telemetry code first retrieves the Starlink two-line element (TLE) data from the Celestrak\footnote{\href{https://celestrak.com/NORAD/elements/supplemental/starlink.txt}{https://celestrak.com/NORAD/elements/supplemental/starlink.txt}} website. Then, using the coordinates of the observatory, it calculates the ephemerides of the satellite and of the Sun. Table\,\ref{Tab.1_new} gives an observing log of the observations presented in this work, including the test observation.

\begin{table} \centering
\caption{\label{Tab.1_new} Log of observations presented for STARLINK-1130 (Darksat) and STARLINK-1113.}
\setlength{\tabcolsep}{3pt} \vspace{-5pt} \begin{small}
\begin{tabular}{lcccc} 
\hline\hline
 & 1130 & 1130  & 1113 \\
 & Darksat & Darksat  & \\
 \hline
 Date (J2000) & 2020/02/08 & 2020/03/06 & 2020/03/06\\
 Time (UTC) & 00:52:30& 00:30:22& 00:15:26 \\
 Filter     & Sloan r'& Sloan g'& Sloan g'\\
 Exposure time (s)& 8.0 & 1.5 & 2.0 \\
 Altitude (km)& 471.62& 563.48&563.97\\
 Range (km)& 516.73&976.50& 941.62\\
 Azimuth ($^\circ$)& 314.37&236.12&215.69\\
 Elevation ($^\circ$)& 64.95&31.48&33.19\\
 Airmass &   1.10         & 1.90    &  1.81   & \\
 RA (Sat)& 04 12 22.71&02 13 36.07&02 09 38.66 \\
 DEC (Sat)&-05 42 21.46&-40 24 55.62&-57 39 01.94   \\
 RA (Sun)& 21 24 34.99& 23 08 06.44&  23 08 04.13 \\
 DEC (Sun)& -15 12 49.45& -05 33 30.31& -05 33 44.79\\
 Sun--Zenith angle ($^\circ$)& 109.2& 110.0& 106.7\\
 Sun--Satellite angle ($^\circ$)& 83.7& 54.7&64.3\\
\hline \end{tabular} \end{small}
\end{table}

\subsection{2020/02/08 test observation}

An initial test observation was performed on 2020/02/08, to determine the accuracy of our telemetry and positional code for Starlink satellites and the effectiveness of our observing technique.

With these ephemerides, we were able to preset the telescope pointing to the required coordinates prior to the arrival of the satellite. To give a high precision measurement of the Darksat trail in the acquired test image, no binning of the detector was used. The readout time for a full-frame, non-binned image of the FLI ProLine 16801 camera is $\approx$\,11.2\,s. We calculated that Darksat had an sky-projected velocity of $\approx1350$\,arcsecs\,s$^{-1}$. For a 11.2\,s readout time, Darksat would end up traversing over 15\,000\,arcsec (over 32\,000\,pixels). Therefore, to reduce the likelihood of Darksat passing through the field of view during readout, the exposure time was set at 8\,s, minimising the dead-time ratio. After the successful test observation, it was decided for future observations to reduce the exposure time in an attempt to capture the full satellite trail within the field of view. 

\subsection{2020/03/06 main observations}

The observation of Darksat conducted on 2020/03/06 was the first observation, during which Darksat had reached both its nominal orbital height (550\,km) and orientation. Therefore, it is not straightforward to directly compare the results from the two observations since once the Starlink satellite has reached its nominal orbit, its orientation changes and it becomes significantly less visible from the ground (SpaceX press kit, \href{https://www.spacex.com/sites/spacex/files/starlink_media_kit_jan2020.pdf}{January 2020}).

Once a Starlink satellite had completed its manoeuvre to its nominal orbit, the TLE data showed that at the apex of the Starlink satellite's trajectory, the sky-projected velocity of the satellite was on the order of a few degrees per second. In an attempt to obtain a full satellite track wholly within the field of view, the observing strategy used in the test observation would need to be changed when using a field of view of $32.4\times 32.4$\,arcmin combined with a readout time of 11.2\,s. To help obtain a full satellite track within the field of view, three adjustments to the observing strategy were undertaken. First, to reduce the sky-projected velocity of the satellite, a lower elevation was selected for the observations. The second change was to reduce the readout time (down to 1.7\,s) by using 4X4 binning of the detector. Thirdly, the exposure times (see Table\,\ref{Tab.1_new}) were selected based on the sky-projected velocity of the satellites (1.5\,s for Darksat and 2\,s for STARLINK-1113) to allow the full trail to be shorter than the field of view. These changes were partially successful with the Darksat observation. However, knowing the predicted ephemerides, even if the track is not fully enclosed in the detector, makes it possible for the satellite's magnitude to be estimated. To give the best approximation to the visual band (550\,nm), the observations of both Darksat and STARLINK-1113 conducted on 2020/03/06 used a Sloan g' filter (475.4\,nm, FWHM 138.7\,nm). 

\subsection{Data reduction}

The raw Flexible Image Transport System (FITS) files were calibrated by removing the instrumental signature from the images by subtracting the bias and dividing by the flat-field. The flux of the comparison stars was measured using standard aperture photometry.

\begin{figure} \includegraphics[width=0.48\textwidth,angle=0]{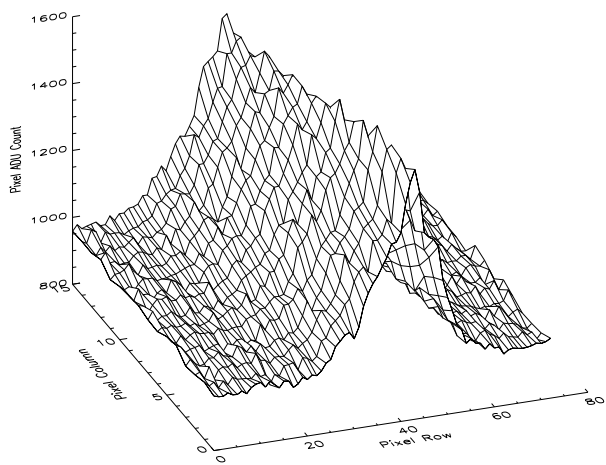} 
\caption{\label{Fig:1}Section of the Darksat trail from the test observation, obtained on 2020/02/08. The section shows a peak pixel count of $\approx$\,1600 and a satellite trail PSF width of 24 pixels. The section is from $x\left(3678:3750\right), y\left(1815:1830\right)$ of the main FITS. } \end{figure}

In examining a section of the Darksat satellite trail from the test observation, it becomes clear that there is a sharp cut-off between the trail point spread function (PSF) and sky background along the trail length (see Fig.\,\ref{Fig:1}). From this we set two parallel boundary lines along the trail length and calculated the total integrated flux after subtracting the sky background.

The integrated flux of the satellite trail was then converted into magnitude using the calibration obtained from the comparison stars.

%--------------------------------------------------------------------
\section{Data Analysis}\label{Sec:3}

The test observation (2020/02/08) contained three suitable comparison stars. However, a review of the literature revealed that the magnitudes had only been determined in the Johnson passbands (Tycho-2 catalogue: \citealt{Tycho2}; Second U.S. Naval Observatory CCD Astrograph Catalog: \citealt{Zacharias2004}). To convert the Johnson magnitudes to the Sloan r' passband, we used the colour equation: $r' = V - 0.42(B-V) +0.11 $ \citep[Table\,1:][]{Jester2005}. The comparison stars in the Darksat and STARLINK-1113 observations conducted on 2020/03/06 have reported Sloan g' magnitudes \citep{Munari2014}, which allowed a direct calculation of the Darksat and STARLINK-1113 apparent magnitudes in the Sloan g' passband. The details of the comparison stars used in this work are given in Table\,\ref{Tab.3}. 

%The results from the three comparison stars show good agreement within their 1-$\sigma$ uncertainty.

\begin{table} \centering
\caption{\label{Tab.3} List of comparison stars used in this work.}
\setlength{\tabcolsep}{2pt} \vspace{-5pt} %\begin{small}
\begin{tabular}{lccccc} 
\hline\hline
Star& Comparison &  Sloan r'   & Sloan g'   \\   
Name &  Star     &  Mag. & Mag. &\\
\hline
 \multicolumn{5}{c}{STARLINK-1130: 2020/02/08} \\
\hline
BD-05\,848 & 1 &  $9.18\pm0.07$ & --  \\
HD\,26528 & 2&  $9.27\pm0.04$ & -- \\
V$^*$ BZ Eri & 3&  $9.67\pm0.06$ & -- \\
\hline
 \multicolumn{5}{c}{STARLINK-1130: 2020/03/06} \\
\hline
CD-40\,570 & 1&  -- & $10.92\pm0.01$  \\
CD-40\,571 & 2&  -- & $10.74\pm0.04$  \\
\hline
 \multicolumn{5}{c}{STARLINK-1113: 2020/03/06} \\
\hline
TYC\,8489-723-1 & 1&  -- & $11.62\pm0.01$  \\
TYC\,8489-699-1  & 2&  -- & $11.90\pm0.01$  \\
\hline \end{tabular} %\end{small}
\end{table}

\begin{figure*} \includegraphics[width=0.33\textwidth,angle=0]{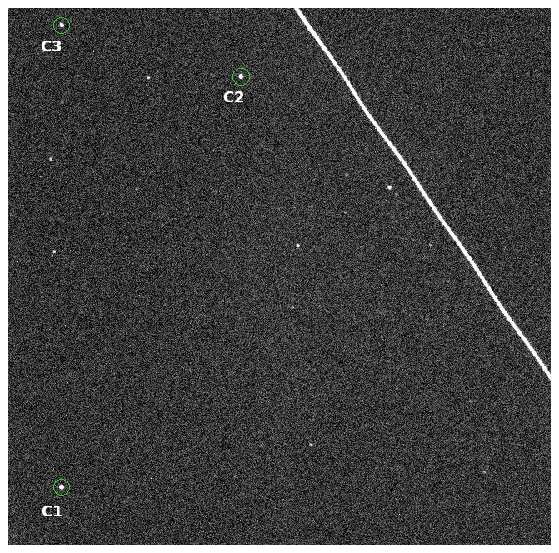} \includegraphics[width=0.33\textwidth,angle=0]{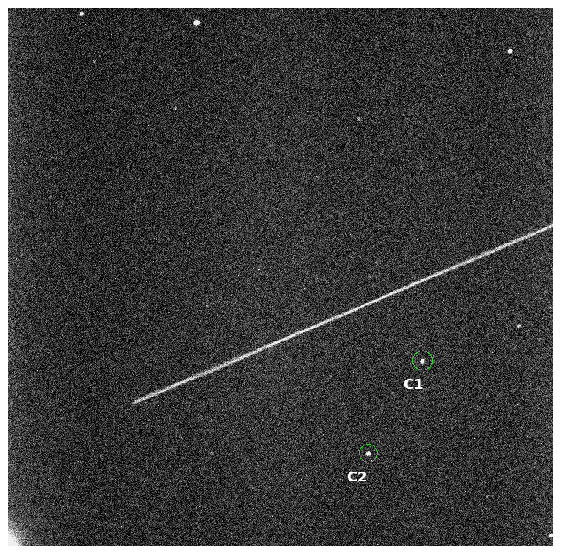} 
\includegraphics[width=0.33\textwidth,angle=0]{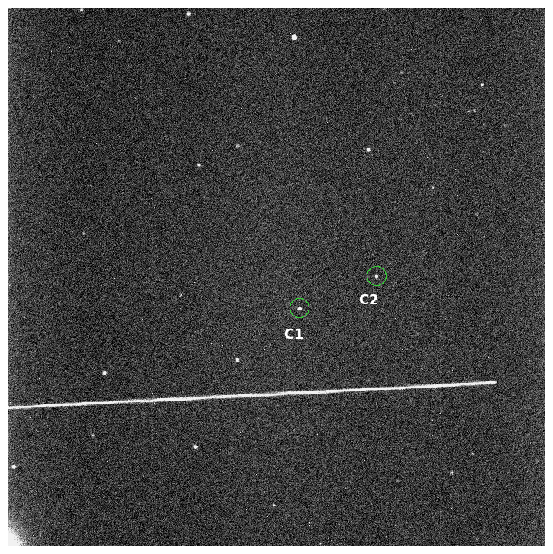} 
\caption{\label{Fig:2}FITS images of STARLINK-1113 and the two Darksat trails presented in this work. The comparison stars are numbered according to Table\,\ref{Tab.3}. {\it left} Darksat: 2020/02/08. {\it middle} Darksat: 2020/03/06. {\it right} STARLINK-1113: 2020/03/06. } \end{figure*}

Figure\,\ref{Fig:2} shows the FITS images of the observations presented in this work, with the comparison stars listed in Table\,\ref{Tab.3} labelled accordingly. Since the Starlink satellite trails in the observations were not wholly within the field of view, only an upper and estimated magnitude could be determined from the observations. The upper magnitude limit was calculated using the measured trail lengths. This leads to a fainter limit of the magnitude for the Starlink satellites. The estimated magnitude was determined by comparing the measured and predicted ephemerides trail lengths, giving an ephemerides-estimated magnitude. The results are presented in Table\,\ref{Tab.4}.

For the Darksat observation conducted on 2020/03/06, the measured and predicted ephemerides trail lengths are in agreement within their 1-$\sigma$ uncertainties, indicating that we successfully observed the full trail length in the image. However, when inspecting the trail in the image, the trail ends a pixel width from the CCD edge, making it difficult to confirm this. Therefore, we provide both the estimated and fainter limit magnitude of Darksat.

\begin{table*} \centering
\caption{\label{Tab.4} Results of the observations presented in this work for Darksat and STARLINK-1113. The results in bold are the weighted means from the individual results from each comparison star.}
\setlength{\tabcolsep}{8pt} \vspace{-5pt} 
\begin{tabular}{lcccc} 
\hline\hline
 Star & Lower  & Estimated & Fainter & Estimated   \\   
      & Diff Mag & Diff Mag & r' Mag & r' Mag \\
\hline
 \multicolumn{5}{c}{STARLINK-1130: 2020/02/08} \\
\hline
BD-05\,848  & $-4.88\pm0.08$ & $-6.88\pm0.09$ & $4.30\pm0.11$ & $2.30\pm0.11$ \\
HD\,26528 & $-4.82\pm0.05$ & $-6.82\pm0.07$ & $4.45\pm0.06$ & $2.45\pm0.08$  \\
V$^*$ BZ Eri & $-5.36\pm0.07$ & $-7.36\pm0.09$ & $4.31\pm0.09$ & $2.31\pm0.11$  \\
\hline
          &               &               & $\mathbf{4.39\pm0.05}$ & $\mathbf{2.38\pm0.06}$ \\
          \hline
 Star & Lower  & Estimated & Fainter & Estimated   \\   
      & Diff Mag & Diff Mag & g' Mag & g' Mag \\
\hline
 \multicolumn{5}{c}{STARLINK-1130: 2020/03/06} \\
\hline
CD-40\,570  & $-3.46\pm0.04$ & $-3.47\pm0.04$ & $7.46\pm0.05$ & $7.45\pm0.05$ \\
CD-40\,571 & $-3.27\pm0.06$ & $-3.28\pm0.06$ & $7.48\pm0.10$ & $7.47\pm0.10$  \\
\hline
          &               &               & $\mathbf{7.47\pm0.04}$ & $\mathbf{7.46\pm0.04}$ \\
\hline
 \multicolumn{5}{c}{STARLINK-1113: 2020/03/06} \\
\hline
TYC\,8489-723-1  & $-4.73\pm0.04$ & $-4.91\pm0.06$ & $6.89\pm0.04$ & $6.60\pm0.06$ \\
TYC\,8489-699-1 & $-5.08\pm0.06$ & $-5.26\pm0.08$ & $6.86\pm0.06$ & $6.58\pm0.08$  \\
\hline
          &               &               & $\mathbf{6.88\pm0.04}$ & $\mathbf{6.59\pm0.05}$ \\
\hline
\end{tabular}
\end{table*}

For the test observation, we measured the trail length to be $3412\pm58$\,pixels, while the ephemerides predicted 22\,900\,pixels for an eight second exposure. This leads to a correction by a factor of $\approx$\,6.7, or 2\,mag, as reported in Table\,\ref{Tab.4}. For a comparison to the observations conducted on 2020/03/06 using a Sloan g' filter, we converted the test magnitude from Sloan r' to g'. We used the colour equation, $g'-r' = 1.05(B-V) - 0.23$ from \citet{Fukugita1996}, with a solar, $B-V=0.65$ \citep{Allen1973}. This gives an estimated Sloan g' magnitude of $\approx2.8$ for the test observation.

The calculated magnitudes given in Table\,\ref{Tab.4} for the respective observations each have a different range ($r$) to the Starlink satellite observed. The orbital height ($H_\mathrm{orb}$) becomes the range when the satellite passes at the local zenith to the observer. Therefore, discerning the reflectivity ratio between Darksat and other Starlink satellites requires that the magnitude is normalised to the nominal orbital height, 550\,km scaled using $+5\log(r/550)$. In addition, the magnitude needs to be corrected for the solar and observer phase angles, $\theta$ and $\phi$, respectively. The observer phase angle is the angle between the observer and the unit normal of the Earth facing surface of the satellite and is approximated by: 

\begin{equation}\label{Eq.1}
 \phi = \arcsin \left( \frac{\eta}{H_\mathrm{orb}} \sin\alpha\right) \ ,
\end{equation}

\noindent where $\eta$ is the straight line distance between the observer and the satellite footprint (nadir) and $\alpha$ is the elevation.
The solar and observer phase angles of Darksat and STARLINK-1113 are given in Table\,\ref{Tab.5}.

While the effects from the two phase angles on a complex body like a Starlink satellite is difficult to model precisely, most of the light observed is diffused. Therefore, we can approximate the effect by using a bidirectional reflectance distribution function (BRDF). Without empirical observations of the BRDF for the Starlink satellites, we can only provide an estimated value by using a parametrised BRDF model from \citet{Minnaert1941}. Consequently, we estimate the ratio ($R$) of the solar phase attenuation between Darksat and STARLINK-1113 with:

\begin{equation}\label{Eq.2}
R = \left(\frac{\cos\theta_{1130}\cos\phi_{1130}}{\cos\theta_{1113}\cos\phi_{1113}}\right)^{k-1} \ ,
\end{equation}

\noindent where $k$ is the Minnaert exponent and ranges from 0 to 1 and $k=1$ represents a perfect Lambertian surface.

Using $k=0$ gives $R\approx0.8$ while, if $k=1$ then $R=1$, for the $\theta$ and $\phi$ of the observations on 2020/03/06. When $k=0.5$ \citep[e.g. a dark surface][]{brdf_book}, then $R\approx0.9$, which is in agreement to a first order approximation of the solar phase attenuation for a diffusing sphere, $(1+\cos\theta)/2$ \citep{Hainaut2020}.
If we assume $k=0.5$ then the solar and observer phase angles would make Darksat appear 0.11\,mag darker than STARLINK-1113, prior to any darkening treatment.

In Table\,\ref{Tab.5} we report the estimated magnitudes observed on 2020/03/06 after correction for the solar and observer phase angles, then normalised to a range of 550\,km. We do not include the test observation from 2020/02/08. As mentioned in Section\,\ref{Sec:2}, after the manoeuvring phase, the orientation of Darksat changes. Hence, the two observations of Darksat are of two different surfaces and this would require too many assumptions to give an accurate result.

\begin{table*} \centering
\caption{\label{Tab.5} Estimated magnitude of Darksat and STARLINK-1113 after correction for the solar and observer phase angles, then normalised to a range of 550\,km (one airmass), including the sky-projected angular velocity along with the observed and estimated trail lengths.}
\setlength{\tabcolsep}{6pt} \vspace{-5pt} %\begin{small}
\begin{tabular}{lccccccccc} 
\hline\hline
Starlink&Observed&Solar Phase&Observer Phase&Angular Speed &  Est. Trail Length&Estimated \\  
Satellite  &  Range (km)& Angle ($^\circ$)&Angle ($^\circ$)& (arcsec\,s$^{-1}$)& (arcsec)& Scaled Mag.\\
\hline
1130 (Darksat)  & 976.50& 54.7 &50.6 &$1075\pm29$&  $1615\pm40$ & $6.10\pm0.04$ \\
1113 & 941.62 & 64.3 & 47.2&$1033\pm27$&  $2066\pm45 $ &  $5.33\pm0.05$ \\
\hline \end{tabular} %\end{small}
\end{table*}

The final results given in Table\,\ref{Tab.5} show that at a range of 550\,km and after correction for the solar and observer phase angles, Darksat is dimmer than STARLINK-1113 by $0.77\pm0.05$\,mag in the Sloan g' passband. This indicates that Darksat is twice as dim as STARLINK-1113.

%--------------------------------------------------------------------
\section{Summary and Discussion}\label{Sec:4}

The successful test observation of the Starlink's Darksat confirm that the ephemerides computed from the publicly available TLE are accurate enough in position and timing to acquire satellite tracks with a professional telescope.

The first observation conducted on 2020/02/08 confirmed that the Starlink satellites are extremely bright ($\approx$2.4\,mag, see Table\,\ref{Tab.4}) during the deployment phase (at 474\,km orbital height). The magnitude was well within the naked-eye sensitivity of a casual observer. During the deployment phase, the Starlink satellites were in groups of 60 and form a train. It is expected that up to three or four satellite trains will be seen in any single night\footnote{\href{https://www.nasaspaceflight.com/2020/01/spacex-launch-third-operational-starlink-mission/}{nasapaceflight.com} accessed on 2020/04/11.} and will, therefore, have an impact on the natural darkness of the sky (see \href{https://www.darksky.org/why-do-mega-constellations-matter-to-the-dark-sky-community/}{International Dark sky Association Press Release}\,; \citealt{Gallozzi2020}). 

The results from the observation of Darksat conducted on 2020/03/06 gives an estimated magnitude in the Sloan g' passband of $7.46\pm0.04$ at a range of 976.50\,km. The observation of STARLINK-1113 conducted on 2020/03/06 gives an estimated magnitude in the Sloan g' passband of $6.59\pm0.05$ at a range of  941.62\,km. This measurement at an elevation of 33.19$^\circ$ validates the photometric model used by \citet{Hainaut2020}, which predicts magnitude 6.6 to 6.7 in these conditions. When scaled to a range of 550\,km and to an elevation of 90.0$^\circ$, our result shows that the brightness of STARLINK-1113 is closer to their fainter estimations. To ascertain a ratio between the reflective flux between Darksat and STARLINK-1113, the magnitude must be corrected for the solar and observer phase angles and for the range normalised to the orbital height or one airmass. This shows that Darksat is $\approx2$ times dimmer than STARLINK-1113. This value should be treated with caution, however. We report an estimated value from using a parametrised BRDF model from \citet{Minnaert1941} and setting the Minnaert exponent to that of a dark surface (i.e. $k=0.5$). To obtain an accurate BRDF measurement will require multiple observations of Darksat and other Starlink satellites during a single trajectory path. Only then  can different empirical reflectance BRDF models (e.g. Phong BRDF: \citealt{Phong1975}; Lewis BRDF: \citealt{Lewis1994}) be compared to the empirical measurements of Darksat and other Starlink satellites.

The results from this work show that Darksat is invisible to the naked eye, even under optimal conditions. However, this reduction does not meet the requirement needed to mitigate the effects that low orbital mega-constellation LEO communication satellites will have on ultra-wide imaging exposures from large telescopes, such as the National Science Foundation’s Vera C. Rubin Observatory (formerly known as LSST). To help mitigate the impact from electronic ghosts in  ultra-wide imaging exposures would require a satellite to be 15 times dimmer than a standard Starlink LEO communication satellite, which would approximately reach down to the 8$^{th}$\,magnitude (see \href{https://www.lsst.org/}{LSST Statement}). However, Darksat is the first response on the part of Starlink with regard to the impacts of mega-constellations of LEO communication satellites on both amateur and professional astronomy. From what we understand, Starlink is studying other methods to decrease the brightness of its satellites, which may  hopefully be deployed in an upcoming launch.

The observations presented in this work, taken when Darksat had reached its nominal orbit (2020/03/06) and orientation, are from a single passband (Sloan g'). At the time of writing, further observations had just been completed at the Chakana 0.6\,m telescope at Universidad de Antofagasta’s Ckoirama observatory, in Sloan r' and i'; whilst on the same nights, observations of Darksat and STARLINK-1113 were conducted in the NIR (J and Ks bands) using VIRCAM (VISTA InfraRed CAMera) on the 4.0\,m VISTA telescope, ESO Paranal, Chile. It is envisaged that the new data will provide a measurement of the change in reflectivity of Darksat as a function of wavelength, from the optical to NIR. Subsequent observations are planned for the period when Darksat is once again visible from northern Chile, including observations aimed  at estimating the BRDF of Darksat and other Starlink satellites.

\begin{acknowledgements}
      We would like to thank the two referees, Dr. Gregg Wade and Dr. Olivier Hainaut, for their helpful comments which improved the quality of this work. This work was supported by a CONICYT / FONDECYT Postdoctoral research grant, project number: 3180071. JTR thanks the Centro de Astronom\'{i}a (CITEVA), Universidad de Antofagasta for hosting the CONICYT / FONDECYT 2018 Postdoctoral research grant. EU kindly acknowledges the work of Marco Rocchetto and Stephen Fossey to set up Ckoirama. We are grateful to both Dr. Patrick Seitzer and Dr. Tony Tyson for useful comments on the manuscript and their encouragement. We extend a special thanks to Boris Haeussler, Felipe Gaete, Steffen Mieske, St\'{e}phane Brilliant, Joseph Anderson, ESO Paranal, for observations which will be included in a follow-up manuscript. The following internet-based resources were used in the research for this paper: the NASA Astrophysics Data System; the ESO Online Digitized Sky Survey, the SIMBAD database and VizieR catalogue access tool operated at CDS, Strasbourg, France; and the ar$\chi$iv scientific paper preprint service operated by Cornell University.
\end{acknowledgements}

\end{document}